\documentclass[10pt,a4paper,twocolumn,american,twocolumn,english,aps,prl,superscriptaddress]{revtex4-1}
\usepackage{lmodern}
\usepackage{lmodern}
\usepackage[T1]{fontenc}
\usepackage[latin9]{inputenc}
\usepackage{color}
\usepackage{amsmath}
\usepackage{amssymb}
\usepackage{graphicx}
\usepackage{esint}

\makeatletter

\@ifundefined{textcolor}{}
{%
 \definecolor{BLACK}{gray}{0}
 \definecolor{WHITE}{gray}{1}
 \definecolor{RED}{rgb}{1,0,0}
 \definecolor{GREEN}{rgb}{0,1,0}
 \definecolor{BLUE}{rgb}{0,0,1}
 \definecolor{CYAN}{cmyk}{1,0,0,0}
 \definecolor{MAGENTA}{cmyk}{0,1,0,0}
 \definecolor{YELLOW}{cmyk}{0,0,1,0}
}

\usepackage{slashed}

\usepackage{babel}

\makeatother

\usepackage{babel}
\begin{document}

\title{Disordered double Weyl node: Comparison of transport and density-of-states calculations}

\author{Bj\"orn Sbierski}
\affiliation{Dahlem Center for Complex Quantum Systems and Institut f\"ur Theoretische
Physik, Freie Universit\"at Berlin, D-14195, Berlin, Germany}

\author{Maximilian Trescher}
\affiliation{Dahlem Center for Complex Quantum Systems and Institut f\"ur Theoretische
Physik, Freie Universit\"at Berlin, D-14195, Berlin, Germany}

\author{Emil J. Bergholtz}
\affiliation{Dahlem Center for Complex Quantum Systems and Institut f\"ur Theoretische
Physik, Freie Universit\"at Berlin, D-14195, Berlin, Germany}
\affiliation{Stockholm University, Department of Physics, SE-106 91 Stockholm, Sweden}

\author{Piet W. Brouwer}
\affiliation{Dahlem Center for Complex Quantum Systems and Institut f\"ur Theoretische
Physik, Freie Universit\"at Berlin, D-14195, Berlin, Germany}

\date{\today}
\begin{abstract}
{\footnotesize{}Double Weyl nodes are topologically protected band
crossing points which carry chiral charge $\pm2$. They are stabilized
by $C_{4}$ point group symmetry and are predicted to occur in $\mathrm{SrSi_{2}}$
or $\mathrm{HgCr_{2}Se_{4}}$. We study their stability and physical
properties in the presence of a disorder potential. We investigate the
density of states and the quantum transport properties at the nodal
point. We find that, in contrast to their counterparts with unit chiral
charge, double Weyl nodes are unstable to any finite amount of disorder
and give rise to a diffusive phase, in agreement with predictions of
Goswami and Nevidomskyy [Phys. Rev. B 92, 214504 (2015)] and Bera, Sau,
and Roy [Phys. Rev. B 93, 201302(R) (2016)]. However, for finite system
sizes a crossover between pseudodiffusive and diffusive quantum 
transport can be observed.}{\footnotesize \par}
\end{abstract}
\maketitle

\section{Introduction}

Topological metals and semimetals are among the driving themes in
contemporary condensed matter physics. Their most prominent three-dimensional
realizations are Weyl (semi)metals, which have recently been experimentally
confirmed in a number of different material systems \cite{Bernevig2015,Xu2015,Lv2015b,Xu2015d,Borisenko2015,Liu2015a}.
Pioneering experimental studies used spectroscopic measurements to
study surface Fermi arcs and characteristic Weyl node bulk dispersions.
Recently, also (magneto-)transport properties received growing interest
in experiment \cite{Zhang2015,Zhang2015a,Shekhar2015}.

While sample quality matures continuously, more controlled engineering
of the chemical potential comes into reach \cite{Ruan2016, Ruan2016a}. Weyl
nodes with chemical potential $\mu$ sufficiently close to the nodal
point ($\mu=0$) are predicted to show unusual transport characteristics
for sample length $L\lesssim\hbar v/\mu$, with $v$ the Fermi velocity
\cite{Baireuther2014,Sbierski2014a}. Without
disorder, the conductance scales with system size as $G\propto W^{2}/L^{2}$,
where $W$ is the sample width. The inclusion of weak disorder is
irrelevant in the renormalization-group (RG) sense \cite{Fradkin1986,Goswami2011,Kobayashi2014,Ominato2014,Syzranov2015a,Syzranov2015cc} and consequently
does not change the size dependence of the conductance. This so-called
``pseudoballistic'' regime is further characterized by an unusual
Fano factor (the ratio of shot-noise power and average current) $F \approx 0.57$
\cite{Baireuther2014,Sbierski2014a}. Only if disorder increases above
a critical strength, the conductivity and density of states 
at the nodal point attains a non-zero value and transport becomes
diffusive, $G\propto W^{2}/L$ and $F=1/3$.

\begin{figure}
\noindent \begin{centering}
\includegraphics{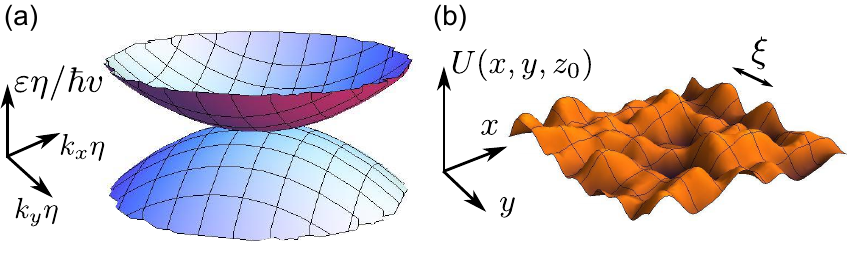} 
\par\end{centering}

\protect\caption{{\small{}\label{fig:summary}(a) Dispersion for DWN Hamiltonian $H$
with $\eta_{x,y}=\eta$ in Eq. (\ref{eq:H}) at $k_{z}=0$. (b) The
potential disorder profile is characterized by its Gaussian correlations
decaying on a length scale $\xi$, here a slice at $z=z_{0}$ is shown.}}
\end{figure}

The simple Weyl node (SWN) band structure discussed above carries
a topological charge of $\pm1$. Beyond the SWN, the existence of
topological band touching points with higher topological charge is
tied to the presence of point-group symmetries \cite{Fang2012a}.
In this paper, we consider double Weyl nodes (DWN) with chiral charge
of magnitude two, stabilized by $C_{4}$ rotation symmetry. The Hamiltonian
reads 
\begin{equation}
H=\hbar v[\sigma_{x}\eta_{x}\,(k_{x}^{2}-k_{y}^{2})/2+\sigma_{y}\eta_{y}\, k_{x}k_{y}+k_{z}\sigma_{z}],\label{eq:H}
\end{equation}
with $\eta_{x,y}$ internal length scales. The fourfold rotational
symmetry around the $z$-axis is realized as $H\!\left(k_{x},k_{y},k_{z}\right)=\sigma_{z}H\left(k_{y},-k_{x},k_{z}\right)\sigma_{z}$.
Time-reversal symmetry is present, $\sigma_{x}H^{*}\left(-\mathbf{k}\right)\sigma_{x}=H\left(\mathbf{k}\right)$
with the time reversal operator $T=\sigma_{x}K$ squaring to $+1$.
To simplify the subsequent analysis, we specialize to the case $\eta_{x}=\eta_{y}\equiv\eta$
where the discrete rotation symmetry is extended to a continuous rotation
symmetry $C_{\infty}$, in cylindrical coordinates $H\left(k_{\perp},\phi,k_{z}\right)=e^{-i\theta\sigma_{z}}H\left(k_{\perp},\phi-\theta,k_{z}\right)e^{i\theta\sigma_{z}}$.
The corresponding energy dispersion $\varepsilon_{\pm}^{2}/(\hbar v)^{2}=(k_{\perp}^{2}\eta/2)^{2}+k_{z}^{2}$
is quadratic in the momentum $\mathbf{k}_{\perp}=(k_{x},k_{y})$ transverse
to the rotation axis and linear in $k_{z}$, see Fig.\ \ref{fig:summary}(a).
A photonic crystal realization of DWNs is reported in Ref. \cite{Chen2015}
and fermionic candidate materials have been identified from first-principle
calculations, such as $\mathrm{HgCr_{2}Se_{4}}$\cite{Xu2011} or
$\mathrm{SrSi_{2}}$\cite{Huang2015}. The latter material might be
experimentally more feasible since no magnetic ordering is required. An interesting proposal to detect the monopole charge in electronic Weyl materials using transport measurements has recently been formulated in Ref. \onlinecite{Dai2016}.

In view of the requirement of point-group symmetries, the stability
of a DWN to disorder, which typically breaks such symmetry {[}see
Fig. \ref{fig:summary}(b){]}, is a relevant question. Several groups
have addressed this question theoretically, with partially diverging
results. Using a simplified version of the self-consistent Born approximation,
Goswami and Nevidomskyy \cite{Goswami2014} argued that the DWN is
unstable to disorder, and that inclusion of even a small amount of
disorder drives the system to a diffusive phase with zero-energy scattering
rate $\hbar/\tau\sim e^{-A/K}$, where $K$ is a dimensionless
measure of the disorder strength and $A$ a material-dependent parameter.
The same conclusion was drawn by Bera, Sau and Roy \cite{Bera2015},
based both on an RG analysis {[}which found disorder a marginally
relevant perturbation to Eq. (\ref{eq:H}){]} and a numerical calculation
of the density of states at zero energy, which was claimed to be compatible with
the exponential form proposed above. 

Recently, Shapourian and Hughes \cite{Shapourian2015} revisited the
same problem, conducting a finite-size scaling analysis of the decay
length in the $z$ direction using a transfer-matrix method. Their data
indicates the presence of a critical point at a \emph{finite} disorder
strength (but below the Anderson transition), leading them to conclude
the stability of the DWN phase against weak disorder. A possible scenario
for such an observation would be the splitting of the DWN into two
equally charged SWNs under the influence of disorder, where the latter
individually would indeed feature a critical point. This interesting
scenario and the apparent contradiction between results in the literature
motivated us to revisit the problem of a disordered DWN. 

We first investigate the density of states using the Kernel Polynomial method
and the self-consistent Born approximation (Sec. \ref{sec:DOS}). We discuss the shortcomings of either method and move on to a scattering
matrix-based transport calculation, much better suited to study the physics
right at the nodal point (Sec. \ref{sec:Quantum-transport}).
These combined numerical efforts allow us to put forward the following
interpretation: In the
presence of any finite amount of disorder, the clean DWN fixed point
is unstable and gives rise to a diffusive phase. We find no evidence 
in support of a critical point at finite disorder strength and, accordingly,
of the DWN splitting scenario. However, due to exponentially small scattering
rate, a crossover behavior can be observed in the quantum transport
properties of weakly disordered mesoscopic samples.

\section{\label{sec:DOS}Density of states}
\subsection{Kernel-Polynomial Method}

We start by calculating the density of states in a disordered DWN which we regularize
on a cubic lattice
\begin{eqnarray}
H_{\mathrm{L}}(\mathbf{k}) & \!= & \!\varepsilon_{0}\frac{\eta}{a}\left[\sigma_{x}(\cos ak_{x}-\cos ak_{y})+\sigma_{y}\sin ak_{x}\sin ak_{y}\right]\nonumber \\
 &  & -\varepsilon_{0}\sigma_{z}\cos ak_{z}
\end{eqnarray}
where $\varepsilon_{0}=\hbar v/a$ and $a$ is the lattice constant.
The effective low energy approximation of $H_{\mathrm{L}}$ around
$\varepsilon=0$ consists of four DWNs centered at $k_{z}=\pm\frac{\pi}{2a}$
and $(k_{x},k_{y})=(0,0)$ or $(\frac{\pi}{a},\frac{\pi}{a})$ with
minimal distance $\Delta k=\pi/a$. 
We include a Gaussian disorder
potential $U(\mathbf{r})$ characterized by zero mean and real space
correlations given by 
\begin{equation}
\left\langle U(\mathbf{r})U(\mathbf{r}^{\prime})\right\rangle _{\rm dis}=\frac{K\left(\hbar v\right)^{2}}{\sqrt{2\pi}^{3}\xi^{2}}e^{-|\mathbf{r}-\mathbf{r}^{\prime}|^{2}/2\xi^{2}},\label{eq:dis_correlator}
\end{equation}
where $\xi$ is the correlation length and $K$ the dimensionless
disorder strength. In the following, we use $\xi=\eta/2$ but different
choices do not qualitatively change our conclusions.
To smoothly represent $U(\mathbf{r})$ on the lattice scale, we take $\xi=2.9a$ which suppresses the inter-node scattering rate by a factor $e^{-(\Delta k)^{2}\xi^{2}/2}<10^{-18}$
compared to the intra-node rate, so single node physics (i.e. $H+U$) is realized to a very good approximation. 

\begin{figure}
\noindent \begin{centering}
\includegraphics{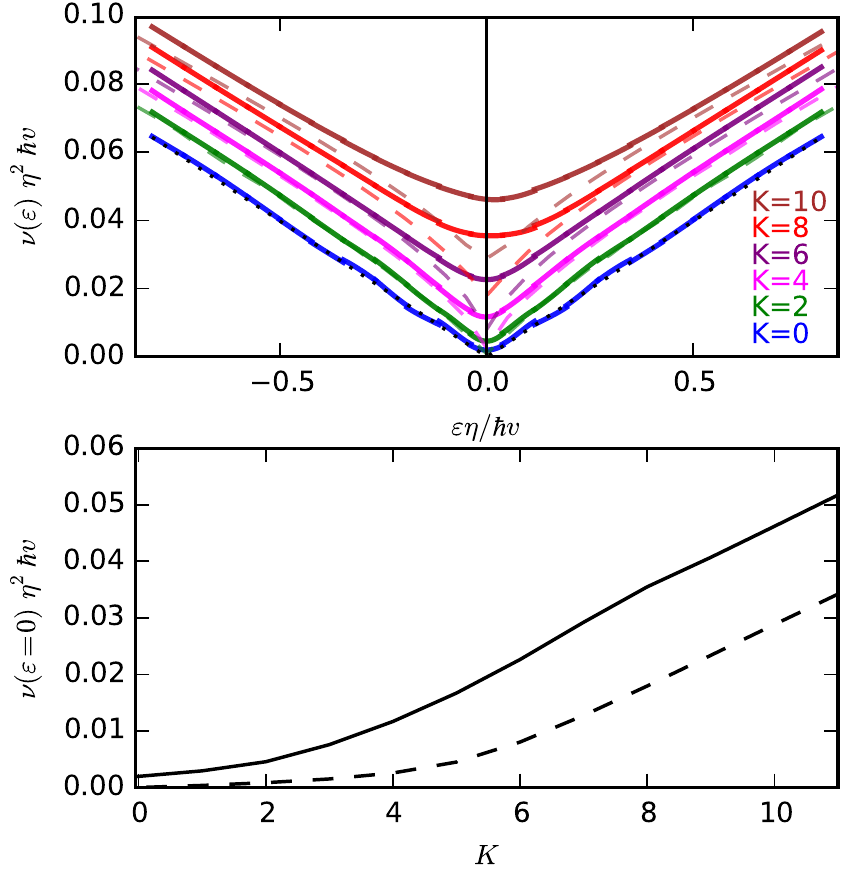}
\par\end{centering}

\noindent \centering{}\protect\caption{{\small{}\label{fig:DOS}Density of states $\nu$ as a function of
energy $\varepsilon$ (top) and at the nodal point $\varepsilon=0$
(bottom) normalized to a}\emph{\small{} }{\small{}single DWN as computed
from the Kernel polynomial method (KPM) applied to the lattice Hamiltonian
$H_{\mathrm{L}}$ (solid lines). The results of the self-consistent
Born approximation based on $H$ (dashed lines) are in good agreement
with the KPM data except in the vicinity of the nodal point and for
large disorder strengths $K>6$. We take the disorder correlation
length $\xi=\eta/2$. The system size of the tight-binding model underlying
the KPM calculation is $L_{x,y}=100a$, $L_{z}=260a$ and we apply periodic
boundary conditions. The expansion order in Chebyshev Polynomials
$N$ is taken in between 1000 and 6000 depending on the energy $\varepsilon$
so that $\nu$ is minimized but oscillations due to the underlying
discrete Eigenenergies of the finite system are sufficiently smoothed
out. An average over 10 disorder realizations is taken and 20 random
vectors were used to calculate the trace in the KPM.}}
\end{figure}

We numerically calculate the density of states of $H_{\mathrm{L}}+U(\mathbf{r})$
using the Kernel Polynomial method (KPM) (see Ref.\ \cite{Weisse2006} for
a description of the method).
The resulting density of states normalized to a single DWN is shown as solid lines
in Fig. \ref{fig:DOS}(a). Further simulation parameters are given in
the figure caption. The analytical result for an infinite clean
system, 
\begin{equation}
\nu_{0}(\varepsilon)=\frac{\varepsilon}{4\pi\left(\hbar v\right)^{2}\eta},\label{eq:cleanDOS}
\end{equation}
is shown as a dotted line in Fig.\ \ref{fig:DOS} and compares well
with the $K=0$ KPM results except at $\varepsilon=0$. At the nodal
point, the KPM method has intrinsic difficulties to simulate the vanishing
(or very small) density of states, which is due to the finite expansion order of
$\nu(\varepsilon)$ in Chebyshev polynomials and the discrete nature
of eigenstates in a finite tight-binding model. In Fig. \ref{fig:DOS}(b),
we plot $\nu(\varepsilon=0)$ vs. $K$. Our findings are in qualitative
agreement with similar numerical results in Ref. \onlinecite{Bera2015}: The
presence of disorder scattering fills the dip in the density of states
for any finite disorder strength.

\subsection{Self-consistent Born approximation}

A frequently employed analytical approach to disordered electronic
systems is the self-consistent Born approximation (SCBA). Although
a simplified SCBA calculation has been performed in Ref. \onlinecite{Goswami2014},
in the following we compute the SCBA self-energy for $H+U$ and the
associated density of states without any further approximations. The
results are shown as dashed lines in Fig. \ref{fig:DOS}, comparison
to the KPM results confirm that SCBA is accurate only at large energies
$\varepsilon$ and weak disorder values, when $\varepsilon\tau\gg\hbar$
with $\tau$ being the quasiparticle scattering time.

We start from Hamiltonian $H$ with $\eta_{x}=\eta_{y}\equiv\eta$
and seek to describe the disorder averaged retarded Green function
$\left\langle G^{R}\right\rangle _{\rm dis}=1/(\varepsilon-H-\Sigma^{R})$
in terms of a translationally invariant self energy term $\Sigma^{R}$
that fulfills the SCBA equation
\begin{equation}
\Sigma^{R}(\mathbf{k})=\int\frac{d\mathbf{k}^{\prime}}{(2\pi)^{3}}\left\langle G^{R}(\mathbf{k}^{\prime})\right\rangle _{\rm dis}\left|U(\mathbf{k}^{\prime}-\mathbf{k})\right|^{2}\label{eq:SCBA equation}
\end{equation}
where $\left|U(\mathbf{k}^{\prime}-\mathbf{k})\right|^{2}$ is the
Fourier transform of the disorder correlator in Eq. \eqref{eq:dis_correlator}.
Since a disorder average restores the $C_{\infty}$-symmetry of the
system around the $k_{z}$ axis, the projection of $\Sigma^{R}(k,\phi,k_{z})$
to the $\sigma_{x}-\sigma_{y}$ plane in Pauli-matrix space should
point into $\phi$ direction, the angle between this plane and the
$\sigma_{z}$ projection of $\Sigma^{R}(k,\phi,k_{z})$ is not dictated
by symmetry and can be different from the angle in $H\left(k,\phi,k_{z}\right)$.
With these considerations, a natural ansatz for the self energy is 
\begin{eqnarray}
\Sigma^{R}(k,\phi,k_{z})/\hbar v & \!= & m(k,k_{z})\!\left(\cos\left[2\phi\right]\sigma_{x}+\sin\left[2\phi\right]\sigma_{y}\right)\nonumber \\
 &  & +\sigma_{z}m_{z}\left(k,k_{z}\right)-im_{0}\left(k,k_{z}\right)\label{eq:SigmaAnsatz}
\end{eqnarray}
with $m$, $m_{z}$ and $m_{0}$ complex and $Re[m_{0}]>0$. At $\varepsilon=0$,
in order to avoid an unphysical spontaneous generation of a
chemical potential from disorder with $\left\langle U(\mathbf{r}))\right\rangle _{\rm dis}=0$,
$m_{0}$ has to be chosen purely real which enforces also $m$ and
$m_{z}$ to be real quantities. The resulting self-consistency equations
for $m$, $m_{z}$ and $m_{0}$ are given in the appendix and can
be solved numerically by iteration. The density of states follows
from
\begin{equation}
\nu\left(\varepsilon\right)=-\frac{1}{\pi}\mathrm{Im}\int\frac{d\mathbf{k}}{(2\pi)^{3}}\mathrm{Tr}\left\langle G^{R}(\mathbf{k})\right\rangle _{\rm dis}.\label{eq:SCBA_DOS}
\end{equation}
Results of this calculation are shown as dashed lines in Fig.\ \ref{fig:DOS} for various representative disorder strengths $K$.

\subsection{Discussion}

The SCBA calculation in Sec. \ref{sec:DOS} can be simplified by taking
the disorder correlation length to zero and choosing a finite (half-)bandwidth
$\Lambda\gg\hbar v/\eta$. Then we could define $K$ such that $\left|U(\mathbf{k}^{\prime}-\mathbf{k})\right|^{2}=K\Lambda^{2}\eta^{3}$
and insert this in Eq. \eqref{eq:SCBA equation} at $\varepsilon=0$,
where $\Sigma^{R}\equiv-i\Gamma$ becomes independent of $\mathbf{k}$.
Transforming to an energy integral and using the density of states
\eqref{eq:cleanDOS} along with the assumption $\Gamma\ll\Lambda$,
one finds 
\begin{equation}
\Gamma=\Lambda e^{-A/K}\label{eq:Goswami}
\end{equation}
with $A=2\pi\left(\hbar v/\Lambda\eta\right)^{2}$. This was first
observed by Goswami and Nevidomskyy in Ref. \onlinecite{Goswami2014} and
states that any finite disorder strength gives rise to a finite lifetime
$1/\Gamma$ of quasiparticles and a finite density of states $\nu(\varepsilon=0)\propto\Gamma$
at the nodal point. Our SCBA analysis which
takes into account a more realistic disorder model and infinite bandwidth
confirms the simplified result in Eq. \eqref{eq:Goswami} qualitatively,
see dashed line in Fig. \ref{fig:DOS} bottom panel.

However, it is well known that the SCBA is not reliable around
gapless points, where the smallness of the parameter $k_{F}l$ spoils
the suppression of crossed diagram contributions to the self energy (see,
{\em e.g.} Ref.\ \cite{Sbierski2014a} for a discussion in the context of simple Weyl nodes). Indeed, comparing the non-perturbative KPM
results for $\nu(\varepsilon)$ to the SCBA in Fig. \ref{fig:DOS},
good agreement is achieved away from the nodal point only. At the
nodal point, it is difficult to judge the qualitative validity of
Eq. \eqref{eq:Goswami} based on the KPM results. The reason is that,
for the latter method, finite size and smoothing effects tend to overestimate
$\nu(\varepsilon=0)$. (For example, the KPM method returns a finite value of $\nu(\varepsilon=0)$ even for $K=0$, see Fig. \ref{fig:DOS}, bottom panel.)
In summary, neither numerical nor analytical calculations of the density of states as presented above are conclusive in gauging the qualitative validity of Eq. \eqref{eq:Goswami} against the alternative scenario of a finite critical disorder strenght below which the bulk density of states vanishes. In this situation, we switch to a quantum transport framework which is ideally suited to study the disordered DWN at the nodal point.

\section{\label{sec:Quantum-transport}Quantum transport}
\subsection{Clean case}
We start this section by calculating the conductance and shot noise power 
of a clean mesoscopic
DWN sample of length $L$ and width $W$ coupled to ideal leads, building
on earlier work by Tworzydlo {\em et al.}\
on two-dimensional Dirac nodes \cite{Tworzydlo2006}.
We choose the transport direction as the $z$ direction and place
the chemical potential at the nodal point. We model the leads as highly
doped DWNs, $H_{\mathrm{lead}}=H+V$ with $V\rightarrow\infty$. By
matching wavefunctions at the sample-lead interfaces we calculate
the transmission amplitudes $t_{0}(\mathbf{k}_{\perp})$ and $t_{0}^{\prime}(\mathbf{k}_{\perp})$
and reflection amplitudes $r_{0}(\mathbf{k}_{\perp})$ and $r_{0}^{\prime}(\mathbf{k}_{\perp})$,
where the primed (unprimed) amplitudes refer to electrons incident
from the positive (negative) $z$ direction, 
\begin{eqnarray}
t_{0}=t_{0}^{\prime} & = & 1/\cosh(\eta Lk_{\perp}^{2}/2),\nonumber \\
r_{0}=-r_{0}^{\prime*} & = & ie^{-2i\varphi}\tanh(\eta Lk_{\perp}^{2}/2),\label{eq:S0-1}
\end{eqnarray}
$\mathbf{k}_{\perp} = (k_x,k_y)$ the transverse component of the wavevector, and $\varphi=\arctan(k_{y}/k_{x})$ is the azimuthal angle of incidence.
The associated basis spinors for propagating states in the lead
are $(0,1)^{\mathrm{T}}$ and
$(1,0)^{\mathrm{T}}$ for left- and right-moving modes, respectively.
From the transmission amplitude $t_{0}(\mathbf{k}_{\perp})$ we compute
the clean-limit conductance and Fano factor as $G_{0}=\frac{e^{2}}{h}\mathrm{tr}[t_{0}^{\dagger}t_{0}]$
and $F_{0}=\mathrm{tr}[t_{0}t_{0}^{\dagger}(1-t_{0}t_{0}^{\dagger})]/\mathrm{tr}[t_{0}t_{0}^{\dagger}]$
\cite{NazarovBlanterBook}. Modes with $k_{\perp}\gg(L\eta)^{-1/2}\equiv k_{\perp}^{\star}\left(L\right)$
are strongly suppressed in transmission and the spacing of the quantized
transversal wave vectors in a finite sample is $\Delta k_{\perp}=2\pi/W$.
If $\Delta k_{\perp}\ll k_{\perp}^{*}$, we can compute conductance
and Fano factor analytically by replacing the sum over discrete modes
$\mathbf{k}_{\perp}$ by an integral and find

\begin{eqnarray}
G_{0}\left(W,L\right) & = & \frac{e^{2}}{h}\frac{1}{2\pi\eta}\frac{W^{2}}{L},\label{eq:G0-1}\\
F_{0}\left(W,L\right) & = & 1/3,\label{eq:F0-1}
\end{eqnarray}
which resembles transport in a diffusive conductor with conductivity
$\sigma_{0}=e^{2}/(2\pi h\eta)$. Thus, the clean DWN has pseudodiffusive
transport characteristics --- similar to Dirac electrons in two dimensions
\cite{Tworzydlo2006}.

\subsection{Disordered case}
We extend the scattering matrix approach to include a Gaussian disorder
potential $U(\mathbf{r})$ with correlations as in Eq. \eqref{eq:dis_correlator} and $\xi=\eta/2$ like in the density of states calculation. We compute the
transmission matrix of the disordered DWN $H+U(\mathbf{r})$ by concatenating
the reflection and transmission amplitudes of a thin slice of DWN
without disorder, see Eq. (\ref{eq:S0-1}), with reflection and transmission
matrices of a thin slice with disorder, which can be calculated using
the first-order Born approximation, and repeating this procedure for
many slices. We apply periodic or antiperiodic boundary conditions
in the $x$ and $y$ directions, cutting off the number of transverse
modes to keep the dimensions of the transmission and reflection matrices
finite. We take the mode cutoff large enough and the slice length
thin enough so that the results do not depend on either, and we take
$\Delta k_{\perp}/k_{\perp}^{*}$ small enough that the results do
not depend on the choice of the boundary conditions. A similar method
has been previously applied to study disordered Dirac materials in
two \cite{Bardarson2007,Adam2009} and in three dimensions \cite{Sbierski2015,Sbierski2014a},
and we refer to those references for more details on the numerical
method.

\begin{figure}
\noindent \centering{}\includegraphics{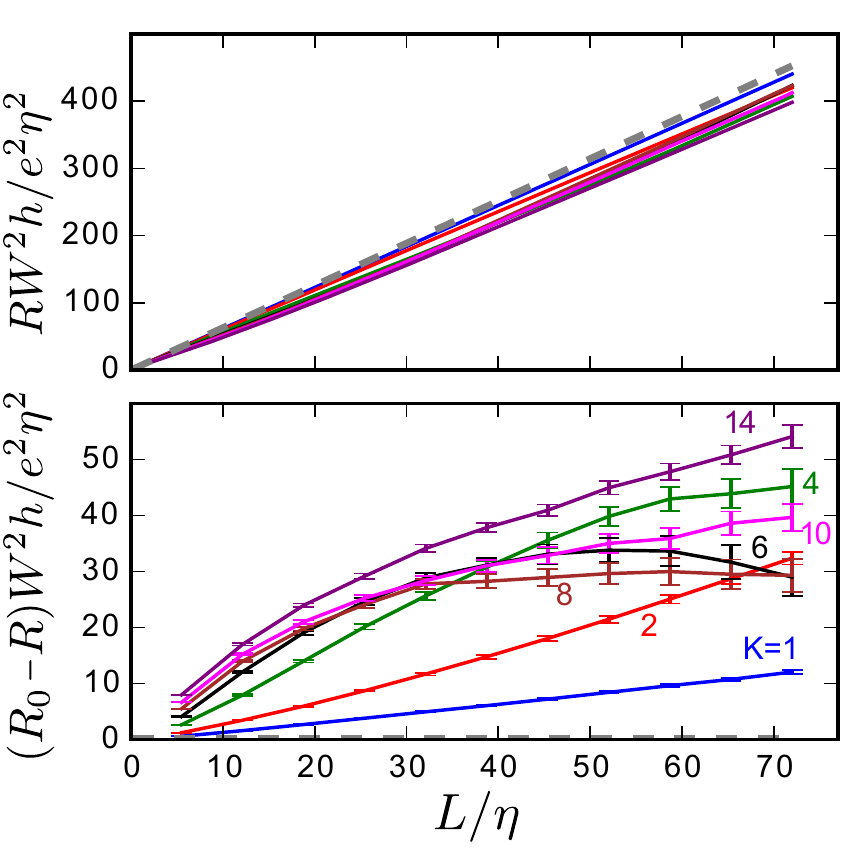}
\caption{{\small{}\label{fig:Transport-properties}Resistance for disordered DWN for disorder strengths $K=1$, $2$, $4$, $6$, $8$, $10$, and $14$ and disorder correlation length $\xi=\eta/2$. Results are averaged over periodic and antiperiodic
transverse boundary conditions and over 60
disorder realizations per boundary condition. The dashed line indicates
the clean-limit pseudodiffusive result of Eq.\ (\ref{eq:G0-1}).
We chose $W=72\eta$, and keep transverse modes with $|k_{x,y}|<2\pi M/W$
with $M=34$. }}
\end{figure}

Figure \ref{fig:Transport-properties} shows our results for the resistance
$R=1/\langle{G}\rangle_{\mathrm{dis}}$ as a function of sample length $L$, where $\langle {...}\rangle_{\mathrm{dis}}$ denotes an average over 60 disorder realizations as well
as the two choices for the boundary conditions, to further suppress
statistical uncertainty. Compared to the clean pseudo-resistance $R_{0}=L/(\sigma_{0}W^{2})$,
the resistance of the disordered samples is slightly decreased by
up to about 10 percent, see top panel. The difference $\Delta R=R_{0}-R$ is 
shown in the bottom panel. For the smallest disorder strength considered,
$K=1$, $\Delta R$ scales linearly with $L$ for the system lengths
considered, for intermediate $K=2,4,6$, $\Delta R$ is not a linear 
function of $L$ but instead has an ``S''-like dependence, 
which prevents any meaningful assignment of a (change of the)
bulk resistivity. The resistance at the largest system size
$R(L_{\rm max}=72\eta)$, shows a non-monotonous behavior with increasing
disorder strength. For larger $K=8,10,14$, the $\Delta R$ traces
are purely convex and tend to be linear for large $L$. We have also
investigated the Fano factor which stays around $F=1/3$ (not shown)
for all values of $K$.

\subsection{Discussion}
\begin{figure}
\noindent \centering{}\includegraphics{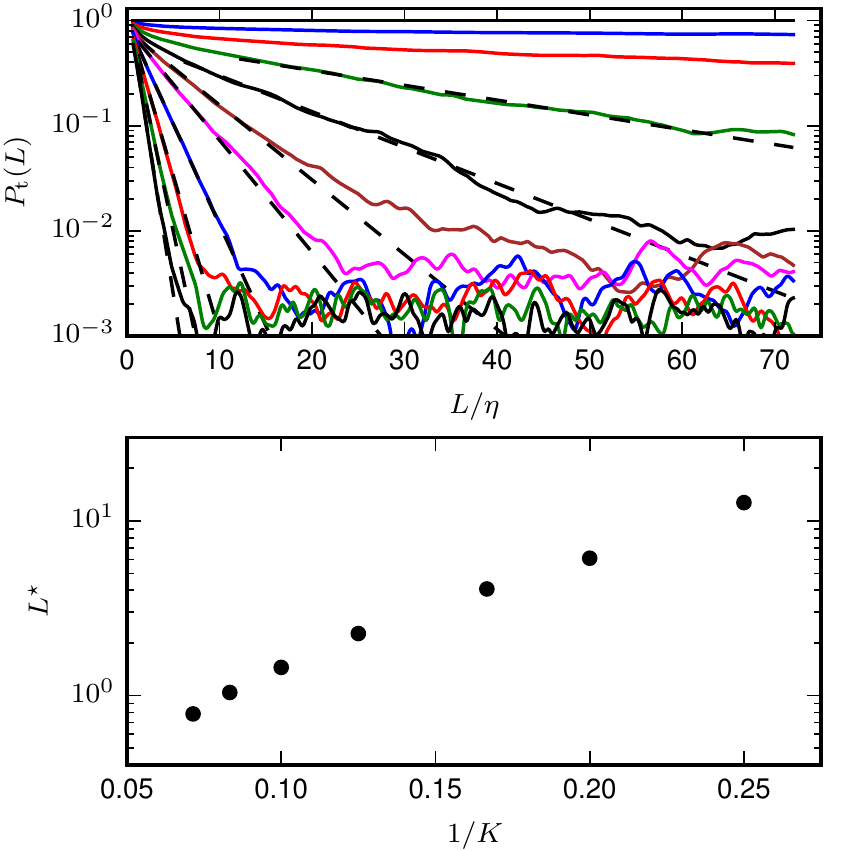}\protect
\caption{{\small{}\label{fig:CrossoverLength} Top: Conditional same-mode transmission probability $P_{\rm t}(L)$ for disorder strengths $K=0, 1$, $2$, $3$, $4$, $5$, $6$, $8$, $10$, $12$, and $14$ (top to bottom curve). The dashed lines denote fits to the form $\exp(-(L-L_{c})/L^{\star})$. Bottom: Mean free path $L^{\star}$, obtained from the fits to $P_{\rm t}(L)$, versus $1/K$.}}
\end{figure}

A finite lifetime $1/\Gamma$ implies diffusive transport with resistance
scaling $R\propto L$. While this is (approximately) observed in our
transport simulations for $K>0$, see Fig. \ref{fig:Transport-properties}
top panel, the difficulty lies in the discrimination to transport
behavior associated to the clean fixed point $K=0$: Being pseudodiffusive,
the same resistance scaling holds, albeit for the very different reason
of evanescent wave physics and not due to scattering between transport
channels as in diffusive transport. To discriminate between the
pseudodiffusive and diffusive regimes, in Fig.\ \ref{fig:CrossoverLength} (top) we show the probability $P_{\rm t}(L)$ that an electron is transmitted in the same transverse mode as it enters --- for which we take the mode with $\mathbf{k}_{\perp} = 0$ ---, conditional on the probability that it is transmitted,
\begin{equation}
  P_{\rm t}(L) = \frac{|t(0,0)|^2}{\sum_{\mathbf{k}_{\perp}} |t(\mathbf{k}_{\perp},0)|^2},
\end{equation}
where $t(\mathbf{k}_{\perp}^{\rm{out}},\mathbf{k}_{\perp}^{\rm{in}})$ is the transmission amplitude of the disordered system at length $L$, $\mathbf{k}_{\perp}^{\rm{out}}$ and $\mathbf{k}_{\perp}^{\rm{in}}$ referring to the incoming and outgoing transverse modes.

The conditional probability $P_{\rm{t}}$ is an indicator of the transition between
the pseudodiffusive and diffusive regimes: At the pseudodiffusive fixed
point $K=0$ one has $P_{\rm{t}}(L) = 1$, as translational translational
invariance ensures that $t(\mathbf{k}_{\perp}^{\rm out},\mathbf{k}_{\perp}^{\rm in})$ is
diagonal in the transversal mode indices $\mathbf{k}_{\perp}^{\rm in}$ and
$\mathbf{k}_{\perp}^{\rm out}$, see \eqref{eq:S0-1}. In contrast, diffusive transport is
characterized by scattering between
transverse modes. For sufficiently long diffusive samples
with many transverse modes one therefore expects $P_{\rm{t}}(L) \to
1/N_{\perp}$, where $N_{\perp}$ is the total number of transverse modes.
For finite-length samples $P_{\rm{t}}(L)$ is expected to approach this
asymptotic value from above, starting from $P_{\rm{t}}(0) = 1$ in the limit of
zero sample length. For the disordered DWN system our data in Fig.\ \ref{fig:CrossoverLength}
(top) indeed indicates a monotonous decrease of $P_{\rm{t}}(L)$ with $L$ and a
saturation at large $L$ for disorder strengths $K > 4$. Although no
saturation could be observed for weaker disorder strength at the system
sizes we could access in our numerical calculations, we found no sign
that $P_{\rm{t}}(L)$ behaves differently for $K < 4$, consistent with with a
flow to a diffusive fixed point even for weak disorder. On the other
hand, if weak disorder is an irrelevant
perturbation (as it is in the case of a single Weyl node) and the
pseudodiffusive fixed point would be stable, we would expect that an
initial decrease of $P_{\rm{t}}(L)$ with $L$ is compensated by increase of
$P_{\rm{t}}(L)$ at larger lenghts, a behavior that we confirmed for the weakly
disordered SWN (data not shown).

As long as $P_{\rm t} \gg 1/N_{\rm t}$, where $N_{\rm t} \sim h G/e^2 \sum_{\mathbf{k}_{\perp}} |t(\mathbf{k}_{\perp},0)|^2$ is the (effective) number of transverse modes participating in the transmission, a condition that is met for the entire parameter range we consider, we expect that $P_{\rm t}(L)$ has the functional form 
\begin{equation}
  P_{\rm t}(L) = e^{-(L-L_{c})/L^{\star}},
\end{equation}
where the characteristic length scale $L^{\star}$ can be identified with the mean free path and $L_{c}$ a length scale that accounts for transient effects at the sample-lead boundary, leading to a quick initial decrease of $P_{\rm t}$ for short lengths, in particular visible for $K\lesssim4$. The lower panel of Fig.\ \ref{fig:CrossoverLength} shows fits of $L^{\star}$ based on the large-$L$ asymptotics of $P_{\rm t}(L)$. The $K$ dependence of $L^{\star}$ is consistent with the expectation based on Eq.\ \eqref{eq:Goswami}, $L^{\star} \sim \hbar v/\Gamma\sim \hbar v/\Lambda e^{A/K}$. We disregard the data points at $K=1$ and $K=2$, for which no reliable asymptotic large-$L$ fit could be made.

The curves for the difference $\Delta R(L)$ of the resistances in the clean and disordered cases in Fig. \ref{fig:Transport-properties} can be understood in terms of a crossover from pseudodiffusive to diffusive transport as well. The length scale $L^{\star}(K)$ roughly coincides with the length scale where the second derivative of the resistance vs.\ sample length curve vanishes.

For the weakest disorder strength we consider the maximum sample length $L_{\rm max}$ is still much smaller than the characteristic length $L^*$ of the pseudodiffusive-to-diffusive crossover. For this disorder strength, pseudodiffusive behavior prevails for all system sizes we consider, albeit with a resistance that is slighlty smaller than $R_{0}$. A decrease of the resistivity has also been observed as a finite-size effect for a SWN at weak disorder strengths \cite{Sbierski2014a}. A systematic decrease of the resistivity could in principle arise as a consequence of a disorder-induced renormalization of the parameters $v$ and $\eta$ in the Hamiltonian (\ref{eq:H}). For a bulk system, the renormalized parameters $v_{\rm eff}$ and $\eta_{\rm eff}$ can be calculated in the Born approximation, which yields an increased effective length scale $\eta_{\rm eff}>\eta$. Replacing $\eta$ by $\eta_{\rm eff}$ in the expression for clean conductivity of a finite system, $\sigma_{0,{\rm eff}}=e^{2}/(2\pi h\eta_{\rm eff})$ predicts an increase of the resistance, in conflict with our numerical observation. We conclude that a disorder-induced renormalization of the parameters $v$ and $\eta$ is not the explanation of the observed decrease of the resistivity. A more careful analysis of the finite-size effects could be attempted along the lines of Ref. \cite{Schuessler2009}.

For strong disorder, $K\gtrsim4$, the characteristic length scale $L^{\star}\left(K\right)$ drops below $L_{\rm max}$ and diffusive behavior can be observed, see, {\em e.g.}, the resistance data for $L/\eta \gtrsim 40$ and $K=14$). Such a diffusive regime is also commonly found in other topological semimetals, such as a two-dimensional Dirac- or a three-dimensional simple Weyl node: Although disorder tends to decrease the mean free path, the conductance is still increased by the disorder-induced increase of the density of states, while band topology and, in three dimensions, standard single-parameter scaling arguments, prohibit Anderson localization \cite{Bardarson2007,Nomura2007,Sbierski2014a}.

\section{Conclusion}

We have investigated the effects of potential disorder for a double
Weyl node, using numerically exact quantum transport simulations
in a mesoscopic setup for chemical potential at the nodal point as well as
density of states calculations based on the self-consistent Born approximation
and the Kernel Polynomial method for a range of energies. Our findings
indicate that disorder physics in a double Weyl node is more
conventional than in its linearly dispersing counterpart with unit
chiral charge, which features a disorder induced quantum phase transition
with the density of states at zero energy as
an order parameter. In the double Weyl node, any finite disorder strength
induces a finite quasiparticle lifetime $\tau$ at the nodal point. 
Our
numerical and analytical calculations are consistent with previous
predictions by Goswami and Nevidomskyy, indicating that the lifetime $\tau$ is exponentially large in the inverse disorder strength \cite{Goswami2014}.

Unfortunately, a quantitative comparison of our calculations for the density of states and our transport simulations is hindered by the fact that only the SCBA can give an estimate for the quasiparticle lifetime $\tau$. However, since the SCBA density of states does not agree quantitatively with the data from KPM at $\varepsilon=0$, we must also discard its predicted value of $\tau$ for quantitative checks. The density of states, as simulated by the KPM is however a quantity integrated over k-space {[}see Eq. \eqref{eq:SCBA_DOS}{]} and cannot be translated into a value for $\tau$ without further assumptions.

In Ref. \cite{Sbierski2015}, the disorder-induced phase transition point in a SWN was identified using the condition of scale invariance of the (median) conductance. We repeated a similar analysis with conductance data obtained for the disordered DWN from Sec. \ref{sec:Quantum-transport} but could not find a scale invariant point (data not shown). This is consistent with the absence of a disorder induced phase transition in a DWN bandstructure.

For technical convenience, we have used a model for the double Weyl node with
continuous rotational symmetry {[}$\eta_{x}=\eta_{y}=\eta$ in Eq.
\eqref{eq:H}{]}. In additional numerical calculations we checked
that our conclusions do not qualitatively change when $\eta_{x}\neq\eta_{y}$
and the rotational symmetry is reduced to be fourfold.

\section*{Acknowledgments}

We gratefully acknowledge discussions with Johannes Reuther and Achim
Rosch as well as financial support by the Helmholtz Virtual Institute
``New states of matter and their excitations'' and the CRC/Transregio 183 (Project A02) of the Deutsche Forschungsgemeinschaft. We thank Jens Dreger
for support on the computations done on the HPC cluster of Fachbereich
Physik at FU Berlin.

\onecolumngrid

\section*{Appendix: SCBA equations}

Using the identities $\int_{0}^{2\pi}d\theta\cos\left(2\theta\right)\exp\left[x\,\text{cos}[\phi-\theta]\right]=2\pi\cos\left(2\phi\right)I_{2}(x)$
and $\int_{0}^{2\pi}d\theta\exp\left[x\,\text{cos}[\phi-\theta]\right]=2\pi I_{0}(x)$
where $I_{k}(x)$ is the modified Bessel function of the $k$-th kind,
we find the following self-consistency equations from Eq. \eqref{eq:SCBA equation}
with the Ansatz \eqref{eq:SigmaAnsatz}:

\begin{eqnarray}
M(P,P_{z}) & = & \frac{-Kr}{\left(2\pi\right)^{2}}\int_{0}^{\infty}dQ\int_{-\infty}^{\infty}dQ_{z}\left[Q^{2}/2+M(Q,Q_{z})\right]I_{2}(QPr^{2})\label{eq:SCBAI}\\
M_{z}(P,P_{z}) & = & \frac{-Kr}{\left(2\pi\right)^{2}}\int_{0}^{\infty}dQ\int_{-\infty}^{\infty}dQ_{z}\left[Q_{z}+M_{z}(Q,Q_{z})\right]I_{0}(QPr^{2})\label{eq:SCBAII}\\
M_{0}(P,P_{z}) & = & \frac{Kr}{\left(2\pi\right)^{2}}\int_{0}^{\infty}dQ\int_{-\infty}^{\infty}dQ_{z}\left[M_{0}(Q,Q_{z})-iE\right]I_{0}(QPr^{2})\label{eq:SCBAIII}
\end{eqnarray}
where 
\begin{eqnarray*}
U(Q,Q_{z}) & = & Q\frac{\exp\left[-r^{2}\left(P^{2}+Q^{2}+(Q_{z}-P_{z})^{2}\right)/2\right]}{[Q^{2}/2+M(Q,Q_{z})^{2}]+[Q_{z}+M_{z}(Q,Q_{z})]^{2}-[E+iM_{0}(Q,Q_{z})]^{2}}
\end{eqnarray*}
and $r=\xi/\eta$, $E/(\hbar v/\eta)=\varepsilon$, $M(Q=q\eta,Q_{z}=q_{z}\eta)\equiv m(q,q_{z})\eta$,
and analogous for $M_{z}$ and $M_{0}$. Eqns. \eqref{eq:SCBAI} to
\eqref{eq:SCBAIII} can be solved numerically by iteration.

\twocolumngrid

\bibliographystyle{apsrev4-1}
\bibliography{library}

\end{document}